\documentstyle[aps,twocolumn,epsf]{revtex}
\begin{document}
\renewcommand{\bottomfraction}{1.0}
\renewcommand{\topfraction}{1.0}
\renewcommand{\textfraction}{0.0}
\renewcommand{\floatpagefraction}{0.9}
\newcommand{\epsplace}[1]{\epsfxsize=3.3in \epsfbox{#1}}

\title
{Trapped Fermi gases}
 
\author{D.A. Butts${}^*$ \and D.S. Rokhsar${}^{*,\dagger}$}
\address
{${}^*$Department of Physics, University of California, Berkeley, CA 94720-7300}
\address
{${}^{\dagger}$Volen Center for Complex Systems, Brandeis University, 
 Waltham, MA 02254}

\address{\em (December 6, 1996)}

\address{ {\em \bigskip \begin{quote} 
We study the properties of a spin-polarized Fermi gas in a harmonic
trap, using the semiclassical (Thomas-Fermi) approximation.  
Universal forms for the spatial and momentum distributions 
are calculated, and the results compared with the corresponding 
properties of a dilute Bose gas.
%
\end{quote} } }
 
\maketitle
\narrowtext

\section{Introduction}
Trapped degenerate 
atomic gases provide exciting opportunities for the manipulation and 
quantitative study of quantum statistical effects, such as the strikingly 
direct observation of Bose-Einstein condensation.\cite{JILA,MIT,Rice}
Although perhaps not as dramatic as the discontinuities associated 
with bosons, the behavior of trapped Fermi gases also merits attention, 
both as a degenerate quantum system in its own right and as a possible 
precursor to a paired Fermi condensate at lower temperatures.\cite{stoof} 
The ideal Fermi gas is an old and well-understood problem; there are many 
familiar systems where the non-interacting Fermi gas is a good zeroth-order 
approximation.  Unlike electrons in atoms and metals and nucleons in
nuclei, however, atomic gases interact by predominantly short-range 
interactions whose effects are weak in the dilute limit.

Trapped atomic gases provide an excellent laboratory for studying 
quantum systems in a controlled, confined setting.  
The quadratic potential of harmonic traps provide a particularly 
simple realization of the confined Fermi system.  At low temperatures,
both the Bose\cite{SGL} and Fermi\cite{Fermi-review} gases are expanded
relative to a classical gas at the same temperature;
for fermions, however, this effect is due to the Pauli exclusion 
principle rather than atom-atom interactions.
While in the Bose case a phase transition separates the degenerate 
and classical regimes, a trapped Fermi gas 
undergoes a gradual crossover between the classical limit and
the compact Fermi sea.

We calculate the spatial and momentum distributions, the chemical potential,
and the specific heat of a trapped spin-polarized Fermi gas,
as a function of temperature.  The properties of gases with two-or-more
trapped spin states will be discussed elsewhere.\cite{dans2}
We find that the properties of harmonically trapped gases with different  
particle numbers can all be described by the same universal 
functions, after suitable scaling of variables.
As we will see below, observation of the spatial distribution of
the trapped cloud would provide an explicit visualization of a 
real-space ``Fermi sea.''

\section{Density of states}
We consider $N$ spin-polarized fermions of mass $M$ moving in an 
azimuthally symmetric harmonic potential, with a single-particle
Hamiltonian
\begin{equation}
{\cal H}({\bf r},{\bf p}) = \frac{1}{2M}[ p_x^2 + p_y^2 +p_z^2 ] +
\frac{M\omega_r^2}{2} [ x^2 + y^2  + \lambda^2 z^2] ,
\label{trap-V}
\end{equation}
where $\omega_r$ and $\omega_z = \lambda \omega_r$ are the trap 
frequencies in the radial and axial directions, respectively.
Since the Pauli exclusion principle prohibits close-approach of
spin-polarized fermions, we may to a first approximation neglect 
short-ranged interactions between them (see below).
The single-particle levels of eq. (\ref{trap-V}) are familiar:
\begin{equation}
\epsilon_{n_x, n_y, n_z} = \hbar\omega_r [ n_x + n_y + \lambda n_z ],
\end{equation}
where $n_x$, $n_y$, and $n_z$ are non-negative integers, and the
zero-point energy has been suppressed.  For temperatures greater
than the level spacing ($k_B T \gg \hbar\omega$), we may treat this
discrete single-particle spectrum as a continuum with density of states
\begin{equation}
g(E) = { {E^2 } \over {2 \lambda (\hbar\omega_r)^3} }.
\label{dos}
\end{equation}

\section{Energy and length scales}
The chemical potential $\mu(T,N)$ is given implicitly by the equation
\begin{equation}
N = \int {{g(E) \, dE} \over {e^{\beta(E-\mu)} + 1}}.
\label{eq:mu}
\end{equation}
At zero temperature the Fermi-Dirac occupation factor is unity
for energies less than the Fermi energy $E_F \equiv \mu(T=0,N)$,
and zero otherwise.  A straightforward integration of eq. (\ref{eq:mu})
then gives\cite{Fermi-review} 
\begin{equation}
E_F = \hbar\omega_r[6\lambda N]^{1/3},
\label{E-fermi}
\end{equation}
which sets the characteristic energy of the atomic cloud.

The characteristic size\cite{Fermi-review} of the trapped degenerate 
Fermi gas is given by $R_F$, the excursion of a classical particle with 
total energy $E_F$ in the trap potential:
\begin{equation}
R_F \equiv [2 E_F/M \omega_r^2]^{1/2} = (48N\lambda)^{1/6} \sigma_{r}.
\label{rf-def}
\end{equation}
where $\sigma_{r} = (\hbar/M\omega_r)^{1/2}$ is the radial width of the
Gaussian ground state of the trap.  For large $N$, the width of the
degenerate Fermi cloud is much greater than the quantum length $\sigma_{r}$,
and the Fermi energy is much greater than the level spacing of the trap,
due to the Pauli exclusion induced ``repulsion'' between 
fermions.\cite{Pauli-potential,Pauli-potential2}

Similarly, we may define a characteristic wavenumber, $K_F$, which
is determined by the momentum of a free particle of energy $E_F$:
\begin{eqnarray}
K_F &\equiv& [{2ME_F}/{\hbar^2}]^{1/2} = (48N\lambda)^{1/6}\; \sigma_{r}^{-1}
\label{kf-def}
\\
&=& (48N\lambda/R_F^3)^{1/3} . 
\label{kf-def2}
\end{eqnarray}
From eq. (\ref{kf-def2}) we see that $K_F$ is roughly the reciprocal 
of the typical interparticle spacing in the gas.

As an example, consider spin-polarized ${}^6$Li. The radial frequency
of this fermionic isotope of lithium in the TOP trap of ref. 
\onlinecite{JILA} would be $\omega_r$ = 3800 sec${}^{-1}$; the
characteristic ground state length $\sigma_{r}$ is 1.6 $\mu$m.
The TOP trap has an intrinsic axial/radial ratio $\lambda = \sqrt{8}$.
For $N=10^5$ atoms, the radius $R_F$ is 25 $\mu$m; the typical interparticle
spacing $1/K_F$ is 0.1 $\mu$m.  The Fermi temperature for this gas would 
be 3.5 $\mu$K.  

\section{Chemical potential and specific heat vs. temperature}
For general temperature, the chemical potential $\mu$ must be determined 
numerically using eq. (\ref{eq:mu}).  We can find analytic expressions, 
however, in the limits of high and low temperature.
For low temperature ($k_B T \ll E_F$) the chemical potential is given by
the Sommerfeld expansion:
\begin{equation}
\mu(T,N) = E_F [1 - \frac{\pi^2}{3}(\frac{k_B T}{E_F})^2] .
\label{Sommerfeld}
\end{equation}
The third and higher-order terms in the Sommerfeld series vanish 
since the density-of-states is a quadratic function of energy.
At high temperatures ({\it i.e.,} in the classical limit 
$k_B T \gg E_F$), we find
\begin{equation}
\mu(T,N) = -k_B T \ln[6( \frac{k_B T}{E_F} )^3] .
\label{classical-mu}
\end{equation}

\begin{figure}
\epsplace{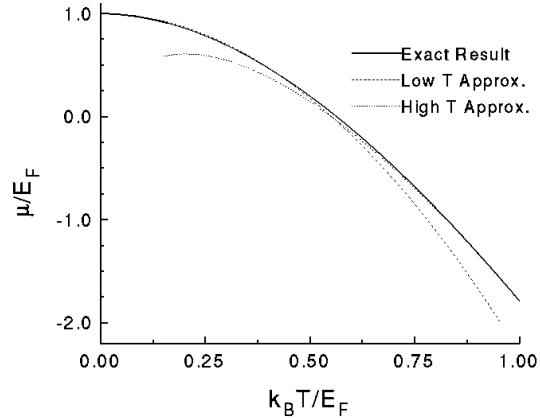}
\caption{The chemical potential {\it vs.} temperature.  Both axes
are scaled by the Fermi energy, which results in a universal curve
that applies to all harmonically trapped Fermi gases.}
\label{fig1}
\end{figure}

Numerical results for $\mu(T)/E_F$ are compared with these two 
limiting forms in figure \ref{fig1}.  Evidently the low temperature
approximation is quantitatively accurate below $k_BT/E_F \sim 0.5$,
and the classical expression holds above $k_B T/E_F$ $\sim 0.6.$
In eqs. (\ref{Sommerfeld}) and (\ref{classical-mu}),
the particle number $N$ enters $\mu(T,N)$ only through the Fermi energy 
$E_F$.  This result holds generally for all temperatures, as can be seen 
by casting eq. (\ref{eq:mu}) in dimensionless form by scaling $E$, $\mu$,
and $1/\beta$ by $E_F$.  (The same conclusion holds for any density
of states of the form $g(E) = A E^b,$ for constant $A$, $b$.)  Figure
\ref{fig1} is therefore a universal curve in the sense that it applies to
harmonically trapped Fermi gases containing any number of particles.

\begin{figure}
\epsplace{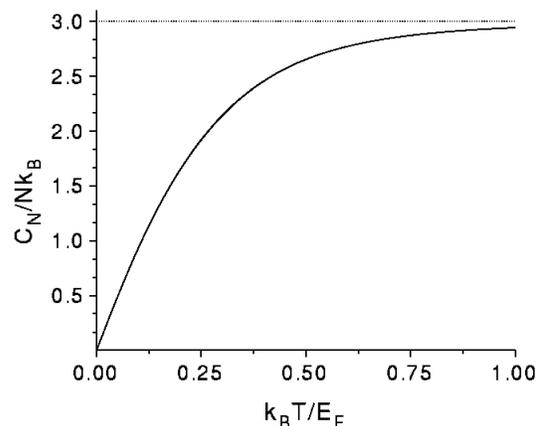}
\caption{The heat capacity {\it vs.} temperature.  The heat capacity
is scaled by $k_B N$ and the temperature by $E_F$.  The classical result
is shown by the dotted line.}
\label{fig2}
\end{figure}

The specific heat\cite{CV-Cp} per particle of the trapped Fermi gas,
$1/N\; \partial U/\partial T |_N$, is shown in figure \ref{fig2}.  
It is a monotonic function of temperature.
For low temperatures the specific heat per particle is
$\pi^2 k_B (k_BT/E_F)$; at high temperatures $C=3Nk_B$,
as we approach the equipartition limit for particles in a 
harmonic potential.

\section{Semiclassical (Thomas-Fermi) approximation}
\label{sec:TF-approx}
Since the exact eigenstates of the harmonic potential are well-known,
the properties of a harmonically trapped ideal gas can in principle
be found directly by summing over these states.
It is useful, however, to have approximate forms for various 
observables that can be computed directly in the large $N$ limit,
where the exact sums become unwieldy.  

In the ``semiclassical'' or Thomas-Fermi approx\-imation,\cite{TF} the 
state of each atom is labeled by a position ${\bf r}$ and a wavevector
${\bf k}$, which can be viewed as the centers of a wavepacket state.
The energy of the particle is simply the corresponding value
of the Hamiltonian; the density of states in the six-dimensional
phase space $({\bf r}, {\bf k})$ is $(2\pi)^{-3}$, where sums over states
are replaced by integrals over phase space.  These semiclassical 
approximations are valid in the limit of large $N$, as discussed 
in the Appendix.

In the semiclassical limit, the number density in phase space is
\begin{equation}
p({\bf r},{\bf k};T,\mu) = \frac{1}{(2\pi)^3}
\frac{1}{e^{\beta({\cal H}({\bf r},\hbar{\bf k})-\mu)} + 1 } .
\label{def-tf}
\end{equation}
The chemical potential is given implicitly by the requirement
\begin{equation}
N = 
\int d^3 {\bf r}\; d^3{\bf k} \ 
p({\bf r},{\bf k};T,\mu) .
\label{eq:TF-mu}
\end{equation}
It follows from the correspondence principle that the Thomas-Fermi
calculation of $\mu(T,N)$ using eq. (\ref{eq:TF-mu}) reproduces the exact 
result obtained from eq. (\ref{eq:mu}); this is easily confirmed for the
harmonic oscillator, since the two integrals are related by a simple 
change-of-variables.

After computing $\mu(T,N)$, it is straightforward to calculate 
the spatial and momentum distribution functions
\begin{eqnarray}
n({\bf r};T) &=& 
\int d^3 {\bf k}\ p({\bf r},{\bf k};T,\mu)  
\label{def-nr}
\\
\tilde{n}({\bf k};T) &=& 
\int d^3 {\bf r}\ p({\bf r},{\bf k};T,\mu)   .
\label{def-nk}
\end{eqnarray}

\section{Spatial distribution at zero temperature}
At zero temperature, we may define a ``local'' Fermi wavenumber 
$k_F({\bf r})$ by
\begin{equation}
\frac{\hbar^2 k_F({\bf r})^2}{2 M} + V({\bf r}) = E_F ,
\label{eq:E-balance}
\end{equation}
where $V({\bf r})$ is the trap potential.
The density $n({\bf r})$ is then simply the volume of the local Fermi sea 
in $k$-space, multiplied by the density of states $(2\pi)^{-3}$:
\begin{equation}
n({\bf r};T=0) = \frac{k_F({\bf r})^3}{6\pi^2}.
\label{eq:k-fermi}
\end{equation}
Note that $n({\bf r})$ vanishes for $\rho > R_F$, where
$\rho$ is the effective distance
\begin{equation}  
\rho \equiv [ x^2 + y^2  + \lambda^2 z^2]^{1/2} .
\label{def-rho}
\end{equation}
Combining eqs. (\ref{eq:E-balance}) and (\ref{eq:k-fermi}), we 
find \cite{Fermi-review,Silvera}
\begin{equation}
n({\bf r};T=0) = \frac{N\lambda}{R_F^3}
\frac{8}{\pi^2} [ 1 - \frac{\rho^2}{R_F^2} ]^{3/2}
\label{eq:ZeroT-distribution}
\end{equation}
for $\rho \leq R_F$.
The cloud encompasses an ellipsoid with diameter $2R_F$ in the $xy$ plane, 
and diameter $2R_F/\lambda$ along the $z$ axis.  
This aspect ratio is the same as that of a classical gas in the same 
potential, since the Boltzmann distribution is proportional to 
$\exp[-M\omega^2\rho^2/2k_B T]$.

\section{Momentum distribution at zero temperature}
One way to characterize the state of a trapped gas is to allow a rapid
adiabatic expansion and then measure the velocity distribution by 
time-of-flight spectroscopy.\cite{JILA}  For the degenerate Bose gas,
the observed anisotropy of this velocity distribution is dramatic
evidence for quantum statistical effects.
The semiclassical momentum distribution for a degenerate Fermi gas at
zero-temperature is simply
\begin{equation}
\tilde{n}({\bf k};T=0) = \frac{1}{(2\pi)^3} \int d^3{\bf r}
\ \Theta( k_F({\bf r}) - |{\bf k}|) ,
\label{k-dist}
\end{equation}
where $\Theta(k_F({\bf r}) - |{\bf k}|)$ is the unit step function.
The integral (\ref{k-dist}) is the real-space volume within which
the local Fermi wavevector exceeds $| { \bf k} |$:
\begin{equation}
\tilde{n}({\bf k};T=0) = \frac{N}{K_F^3}
\frac{8}{\pi^2} [ 1 - \frac{|{\bf k}|^2}{K_F^2} ]^{3/2} ,
\label{eq:k-distribution}
\end{equation}
where the maximum occupied wavenumber $K_F$ was defined in eq. (\ref{kf-def}).
Note that $k_F({\bf r}=0) =$ $K_F$.

Despite the spatial anisotropy of the trap, the momentum distribution 
of the degenerate Fermi gas is isotropic.  This isotropy is a general 
feature of trapped Fermi gases, independent of trap potential,\cite{box}
since from eq. (\ref{k-dist}) we see that $\tilde{n}({\bf k})$ depends 
only on the magnitude of ${\bf k}$.  

The spatial and momentum distributions (\ref{eq:ZeroT-distribution},
\ref{eq:k-distribution}) both have the same functional form, because
${\cal H}$ is a quadratic function of both position and momentum.
In this sense, the distribution (\ref{eq:ZeroT-distribution}) can be 
viewed as a Fermi sea in real-space.  
The anisotropy of $n({\bf r})$ is due to the unequal spring constants, 
while the isotropy of $\tilde{n}({\bf k})$ is due to the isotropy of mass.  
That is, $p_x^2$, $p_y^2,$ and $p_z^2$ enter the Hamiltonian with the
same coefficient, while $x^2,$ $y^2,$ and $z^2$ need not.

\section{Numerical results}
In the semiclassical approximation, the spatial and momentum distributions 
are easily determined numerically for any temperature as described 
in section \ref{sec:TF-approx}.
As with the chemical potential, an appropriate scaling of these two
distributions yields a universal form for all harmonically trapped 
Fermi gases when plotted {\it vs.} the scaled variables $\rho/R_F$ and 
$|{\bf k}|/K_F$, respectively.

\begin{figure}
\epsplace{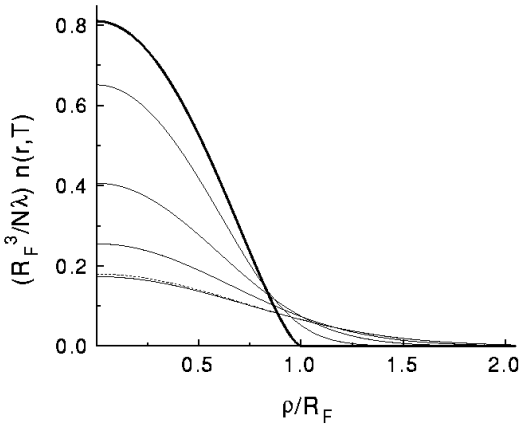}
\caption{The universal spatial and momentum density distributions for 
$k_B T/E_F=0\ $(bold),$\ 0.25$,$\ 0.5$,$\ 0.75$, and $1.0$.  The 
classical result for $k_B T/E_F=1$ is shown as a dashed line.}
\label{fig3}
\end{figure}

Figure \ref{fig3} shows the scaled density {\it vs.} scaled distance
for $k_B T/E_F$ of 0, 0.25, 0.5, 0.75, and 1.   
At low temperatures, the  density is close to their zero temperature
form, with a thin evaporated ``atmosphere'' of thickness 
$\sim R_F (k_BT/E_F)$ surrounding a degenerate liquid ``core.''

\begin{figure}
\epsplace{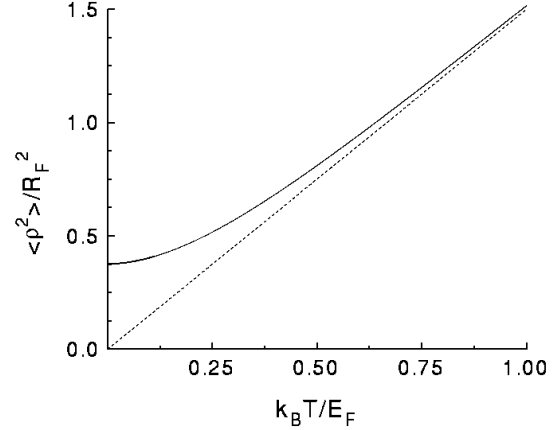}
\caption{The mean-square variation of the size of the cloud
interpolates between a low-temperature degenerate regime and a
high-temperature limit that is well-described by the equipartition
theorem.}
\label{fig4}
\end{figure}

In the classical limit, the density approaches a Gaussian in $\rho$,
with a width given by the equipartition theorem:
$\langle \rho^2 \rangle = 3 R_F^2 (k_B T/E_F).$
As shown in figure \ref{fig3}, this accurately describes the density 
distribution for $k_B T/E_F=1$.
The evolution of the density profile from its low-temperature Fermi
form (\ref{eq:ZeroT-distribution}) to the classical limit 
can be tracked by calculating the mean-square excursion
$\langle \rho^2 \rangle$, which is shown in figure \ref{fig4} in the 
dimensionless form $\langle \rho^2 \rangle/R_F^2$  {\it vs.}
$k_B T/E_F$.  This is again a universal curve for all harmonically
trapped Fermi gases.

As noted above for zero temperature, the momentum distribution at
non-zero temperature has the same form as the spatial distribution,
since momentum and position both enter the single particle Hamiltonian
quadratically.  Thus the scaled momentum distribution 
$K_F^3 \tilde{n}({\bf k})/N$ vs. $|{\bf k}|/K_F$
is also given by figure \ref{fig3}.  Similarly, figure \ref{fig4}
illustrates the scaled mean-square momentum $\langle {\bf k}^2 \rangle
/K_F^2$ {\it vs.} $|{\bf k}|/K_F$ as well.

\section{Perturbations}
What happens if the potential is not perfectly harmonic?  We may treat
$\delta V({\bf r})$ as a perturbation.  Here we focus our attention
on the $T=0$ case.  From eq. (\ref{eq:E-balance}), a change in trap 
potential shifts the local Fermi wavenumber by 
\begin{equation}
\delta k_F({\bf r}) = \frac{M}{\hbar k_F({\bf r})} 
[ \delta E_F - \delta V({\bf r})],
\label{perturb}
\end{equation}
where $\delta E_F$ is the change in Fermi energy.  From eq. 
(\ref{eq:k-fermi}), the corresponding change in density is
\begin{equation}
\delta n({\bf r}) = \frac{M k_F({\bf r})}{2\pi^2\hbar}
[ \delta E_F - \delta V({\bf r})],
\label{delta-n}
\end{equation}
where the Fermi energy is adjusted to make 
$\int d^3{\bf r}\; \delta n({\bf r})$ vanish:  
\begin{equation}
\delta E_F = \frac{\int d^3 {\bf r} \delta V({\bf r}) k_F({\bf r})}
{\int d^3 {\bf r} k_F({\bf r})} .
\end{equation}

Interactions between
particles (which are small for spin-aligned gases) can be treated
as a perturbation $\delta V({\bf r}) = U n({\bf r})$, where $U$
is a parameter related to the $p$-wave scattering length.

\section{Comparison with the Bose gas}
The interacting Bose gases of refs. \onlinecite{JILA} and \onlinecite{MIT}
are in the Thomas-Fermi regime\cite{BP}.
Since the gases remain dilute, two-body scattering 
may be treated by a delta-function pseudopotential of strength 
$U = 4\pi \hbar^2 a/M$, where $a$ is the $s$-wave scattering length.  
When the dimensionless parameter $UN/\hbar\omega\lambda\sigma_{r}^3$ 
is large (as is appropriate for the experiments of refs. \onlinecite{JILA}
and \onlinecite{MIT}) the density profile of the interacting Bose 
gas is \cite{BP}
\begin{equation}
n_B({\bf r}) = {{R_B^2} \over {2U}} \ 
\big[ 1 - \frac{\rho^2}{R_B^2} \big] ,
\label{bose-dist} 
\end{equation}
with maximum radius 
\begin{equation} 
R_B = (\frac{15\lambda UN}{4\pi})^{1/5}.
\label{rb-def}
\end{equation}
Note that the characteristic radius scales more slowly with particle
number for the Fermi gas ($N^{1/6}$) than for the interacting Bose gas
($N^{1/5}$); similarly, the Fermi energy scales as $N^{1/3}$ while
the zero temperature chemical potential of the Bose gas varies more
rapidly, as $N^{2/5}$.

The axial/radial aspect ratio for both classical and degenerate
trapped gases is $\lambda$, since in every case the densities are
functions of $\rho$ only.  The velocity (momentum) distributions,
however, can be quite different: for classical and Fermi gases the 
velocity distribution is isotropic, while $\tilde{n}({\bf k})$
for a zero temperature Bose gas is the square of the Fourier transform of 
$\sqrt{n_B({\bf r})}$.  This is notably anisotropic.
Note that as $N$ increases, both $R_F$ and $R_B$ increase, but
the widths of the respective {\em momentum} distributions
go in opposite directions: $K_F$ increases with $N$, while
the typical momentum of a particle in a trapped Bose condensate
decreases with particle number, since $K_B \sim 1/R_B$ by the 
uncertainty principle.

It is amusing to compare the interatomic repulsion in a Bose gas with 
the effective repulsion experienced by fermions due to the Pauli exclusion
principle.  Equating the characteristic Fermi and Bose radii 
(\ref{rf-def}, \ref{rb-def}) we see that, crudely speaking, the spatial
distribution of a degenerate Fermi gas is mimicked by that of a Bose gas
interacting via an effective ``Pauli pseudopotential'' 
$U_{\rm eff} \sim E_F(R_F^3/N)$, which
is the characteristic energy multiplied by the volume per particle. 
Equivalently, the effective scattering length brought about by the
Pauli principle is $a_{\rm eff} \sim K_F^{-1}$, {\it i.e.,} the
inter-particle spacing.  
(This is quite expected, since the inter-particle
spacing is the only appropriate length in the ideal Fermi gas.)  
The use of such an effective interaction is limited by the fact that 
(a) the momentum distributions of the Fermi and Bose gases remain quite 
different and (b) the gas is not dilute with respect to the 
exclusion-induced ``interactions'' since $K_F a_{\rm eff}$ is of order unity.

\section{Acknowledgements}  We thank Mark Kasevich, David Weiss, and
Ike Silvera for useful and interesting discussions.  This work was supported 
by the National Science Foundation under grant NSF-DMR-91-57414, and the 
Committee on Research at UC Berkeley.  Work at Brandeis was supported by 
the Sloan Center for Theoretical Neurobiology.

\section{Appendix: Validity of the semiclassical approximation}
The semiclassical approximation can be safely applied to an
inhomogeneous Fermi gas of density $n({\bf r})$ if we can imagine
partitioning the system into cells of linear 
dimension ${\ell}$ such that the following two conditions are 
simultaneously met:

(1) The number of particles in a cell is much greater than unity,
so that a local Fermi sea may be envisioned:
\begin{equation}
n(r) {\ell}^3 \gg 1.
\label{appendix-1}
\end{equation}

(2) The variation of the trap potential across the cell 
(${\ell}\nabla V$) must be small
compared with the local Fermi energy $\hbar^2 k_F(r)^2/2M$, so
that within a cell the potential energy is nearly constant.
At low temperature, this condition becomes
\begin{equation}
\ell M\omega^2 r \ll \frac{\hbar^2}{2M} [ 6\pi^2 n(r) ]^{2/3},
\label{appendix-2}
\end{equation}
where we have used eq. (\ref{eq:k-fermi}).

Combining eqs. (\ref{appendix-1}) and (\ref{appendix-2}), we see that
to be able to choose a (possibly $r$-dependent) cell size ${\ell}$
that simultaneously satisfies these two conditions, 
the number of particles per quantum volume must satisfy
\begin{equation}
n(r) \sigma^3 \gg \frac{r}{\sigma},
\label{appendix-3}
\end{equation}
where $\sigma$ is as before the quantum length $(\hbar/M\omega)^{1/2}$
and we have omitted factors of order unity.  

At low temperatures, the semiclassical density given by eq. 
(\ref{eq:ZeroT-distribution}) scales as $N/R_F^3 \sim N^{1/2}/\sigma^3$ 
near the origin, so the Thomas-Fermi approximation is always self-consistent 
at the center of the trap for large $N$.  (This can be confirmed at ${\bf r}=0$
by direct summation of the squares of the simple harmonic oscillator 
eigenfunctions up to energy $E_F$.)
Near the periphery of the cloud, however, the density becomes small,
and the approximation is not valid.  It is easy to show that semiclassical
treatment fails within a shell at the periphery of the cloud,
whose thickness $\delta R \sim$ $1/K_F \sim$ $\sigma N^{-1/6}$
vanishes in the limit of large $N$.  Within
this shell only the exponential tails of a few single particle states
contribute to the density; this is analogous to the corresponding 
region of the Bose gas.\cite{String-Daf}


\begin{references}
\bibitem{CV-Cp}
{Note that the trapped gas does not occupy a container of 
externally fixed size, and so we do not introduce separate
constant-volume and constant-pressure specific heats.
As the gas is heated, it expands against the harmonic potential, 
doing work; the specific heat of the harmonically trapped system 
is therefore expected to be larger than that of a comparable 
gas-in-a-box at fixed volume.  Since the harmonic restoring force
grows linearly with the size of the cloud, the effective pressure 
it exerts will also vary.}
\end{references}
\end{document}